\documentclass[conference, 11pt]{IEEEtran}
\usepackage{fancyvrb}
\IfFileExists{upquote.sty}{\usepackage{upquote}}{}
\usepackage{cite}
\usepackage{amsmath,amssymb,amsfonts}
\usepackage{algorithmic}
\usepackage{graphicx}
\usepackage{textcomp}
\usepackage{xcolor}
\usepackage{subcaption}
\usepackage{url}
\usepackage{array}
\usepackage{listings}
\usepackage{multirow}
\usepackage{tabularx}
\usepackage{booktabs}
\usepackage{siunitx}
\usepackage{listings}
\usepackage[T1]{fontenc}
\usepackage{xcolor} % For custom colors

\lstdefinestyle{promptstyle}{
    backgroundcolor=\color{black!5},   % very light grey background
    basicstyle=\footnotesize\ttfamily, % smaller, monospaced font
    breaklines=true,                  % wrap long lines
    postbreak=\mbox{\textcolor{red}{$\hookrightarrow$}\space}, % symbol for wrapped lines
    frame=single,                     % adds a frame around the code
    framerule=0.4pt,
    rulecolor=\color{black!60},
    framesep=5pt,
    showstringspaces=false,
    tabsize=2
}

\usepackage[top=0.75in, left=0.63in, right=0.63in]{geometry}

\def\BibTeX{{\rm B\kern-.05em{\sc i\kern-.025em b}\kern-.08em
    T\kern-.1667em\lower.7ex\hbox{E}\kern-.125emX}}
\begin{document}

\title{A Decompilation-Driven Framework for Malware Detection with Large Language Models}
\author{\IEEEauthorblockN{Aniesh Chawla\IEEEauthorrefmark{1}, Udbhav Prasad\IEEEauthorrefmark{1}}
\IEEEauthorblockA{\IEEEauthorrefmark{1}
California, USA \\
\{chawla.aniesh, udbhav523\}@gmail.com}
\IEEEauthorblockA{\IEEEauthorrefmark{1}These authors contributed equally to this work.}}

\maketitle

\begin{abstract}
The parallel evolution of Large Language Models (LLMs) with advanced code-understanding capabilities and the increasing sophistication of malware presents a new frontier for cybersecurity research. This paper evaluates the efficacy of state-of-the-art LLMs in classifying executable code as either benign or malicious. We introduce an automated pipeline that first decompiles Windows executable into a C code using Ghidra disassembler and then leverages LLMs to perform the classification. Our evaluation reveals that while standard LLMs show promise, they are not yet robust enough to replace traditional anti-virus software. We demonstrate that a fine-tuned model, trained on curated malware and benign datasets, significantly outperforms its vanilla counterpart. However, the performance of even this specialized model degrades notably when encountering newer malware. This finding demonstrates the critical need for continuous fine-tuning with emerging threats to maintain model effectiveness against the changing coding patterns and behaviors of malicious software.

\end{abstract}


\section{Introduction}

Traditional signature-based malware detection methods increasingly struggle to identify novel malware variants that employ advanced obfuscation techniques, polymorphic transformations, and previously unseen attack vectors. This limitation has sparked significant interest in leveraging artificial intelligence, particularly Large Language Models (LLMs), to enhance malware detection capabilities through advanced code comprehension and semantic analysis \cite{Zhang2025MalwareLLM}\cite{Wu2024BinaryLLM}\cite{Papernot2025CyberAI}.

Large Language Models have demonstrated remarkable proficiency in understanding and analyzing code across various programming languages, making them promising candidates for binary analysis \cite{Chen2025LLMSecurityGraphs} and malware classification tasks \cite{Li2025AIReverseEngineering}. Recent research has explored LLM applications across multiple cybersecurity domains like vulnerability detection, network intrusion analysis, and malicious code identification. However, the direct application of LLMs to decompiled executable code for malware classification represents a relatively underexplored frontier with significant potential for advancing automated threat detection capabilities.

\begin{figure}
    \centering
    \includegraphics[width=0.9\linewidth]{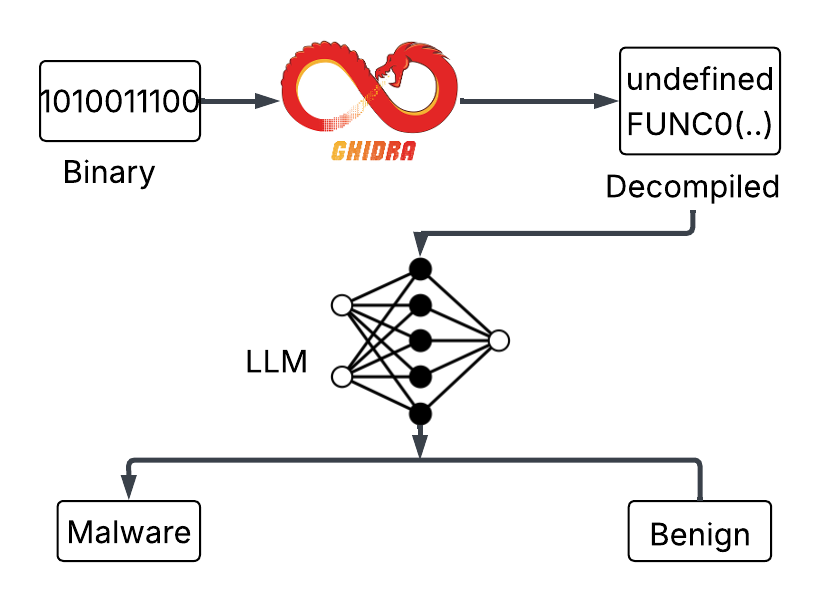}
    \caption{System to classify Malware/Benign binary file}
    \label{fig:overall-system}
\end{figure}

The integration of reverse engineering tools, particularly the Ghidra \cite{ghidra} framework, with modern LLMs presents a novel methodology for automated malware analysis. Ghidra's sophisticated decompilation capabilities can transform binary executables into human-readable C code, creating an intermediate representation that LLMs can effectively process and analyze. This approach bridges the gap between low-level binary analysis and high-level semantic understanding, potentially enabling more nuanced and context-aware malware detection than traditional static analysis methods.

This paper introduces an automated pipeline that combines Ghidra's decompilation capabilities with state-of-the-art LLMs to classify Windows executables as either benign or malicious. Our research addresses critical limitations identified in previous studies, including the reliance on hypothetical datasets and the lack of real-world applicability. By utilizing authentic malware samples from established repositories and contemporary benign executables, we evaluate the practical effectiveness of both vanilla and fine-tuned LLMs in production-relevant scenarios.

The primary contribution of this work lies in demonstrating how fine-tuning significantly enhances LLM performance on malware classification tasks. This work also reveals the limitations that necessitate continuous model updates to maintain effectiveness against emerging threats. Our findings provide crucial insights into the practical deployment challenges and opportunities for LLM-based malware detection systems, contributing to the broader understanding of artificial intelligence applications in cybersecurity defense strategies.
\section{Related Work}

\subsection{Surveys and Systematic Reviews}

Khan et al. (2024) \cite{Khan2024ExploringLLMs} present a comprehensive review of the applications, challenges, and risk mitigation strategies surrounding the use of large language models (LLMs) for malware detection and prevention. Their study analyzes the existing literature, categorizing it into key areas such as harmful content generation, weaponization, malware dissemination, benchmarking, and new model proposals. The paper examines major uses of LLMs, ranging from honeypot generation and text-based threat identification to code analysis, trend detection, and disguised malware recognition. Despite these contributions, the paper is limited by its use of hypothetical rather than real-world data, gaps in research coverage on weaponization and malware distribution, and challenges surrounding the explainability and interpretability of LLM-based decisions. Additionally, the necessity for frequent model retraining poses practical challenges for integration into live security systems.

The approach taken by our paper overcomes these limitations using real malware samples collected from various sources (\cite{saxe2018malware}, \cite{MalwareBazaar}) as well as common and uncommon benign files ensures applicability of the discussed techniques in production scenarios. Moreover, our paper demonstrates the significant positive effects of fine tuning instead of full retraining on custom data, which is of great practical use to large enterprises.

Wang et al. \cite{wang2025contemporary} present a comprehensive survey of LLM applications in program analysis, systematically categorizing existing work into static analysis, dynamic analysis, and hybrid approaches across vulnerability detection, malware detection, program verification, fuzzing, penetration testing, and unit test generation. The authors demonstrate how LLMs enhance traditional program analysis through context-aware code comprehension and semantic understanding, leveraging intermediate representations such as Abstract Syntax Trees (AST), Control Flow Graphs (CFG), and Data Flow Graphs (DFG). However, the paper acknowledges significant limitations ranging from non-deterministic outputs leading to inconsistent results and hallucinations and practical constraints involving high costs, token limits restricting scalability, and heavy reliance on prompt engineering expertise.

\subsection{Reverse Engineering and Binary Analysis}

Pordanesh and Tan \cite{pordanesh2024gpt4} analyze how GPT-4 and similar architectures interpret and explain decompiled or disassembled code. Their experiments show how LLMs facilitate malware analyst workflows, including function inference and behavior annotation from raw binaries. However, their study reveals significant limitations in GPT-4's ability to handle complex control structures and intricate code relationships, with particular weakness in identifying stealth techniques (46.7\% error rate) and establishing logical connections between code components.

Technical reports from industry (e.g., CrowdStrike \cite{crowdstrike2025byteback}) demonstrate the deployment of byte-level transformers directly on binaries for scalable malware classification, often outperforming classical signature-based methods. However, these industry approaches lack detailed methodological transparency and independent validation, making it difficult to assess their effectiveness against sophisticated evasion techniques.

Rondanini et al. \cite{rondanini2024edge} analyze the ability of open-source LLMs to describe file functions extracted from binaries, as well as classifying API calls relevant for threat detection. These studies are constrained by limited context windows that restrict analysis of large binary files and struggles with maintaining accuracy as input complexity increases. While our approach is also limited by the context windows offered by off-the-shelf models, we show how input complexity can be overcome with fine-tuning.
\section{Experiment Setup}
\subsection{Data Collection} \label{sec:datacollection}
To construct a comprehensive dataset for our evaluation, we aggregated executable files from two primary sources: a well-established academic data set and a repository of contemporary, most recent threats.

\subsubsection{Baseline Dataset}
Our baseline dataset is derived from the corpus provided in the \textit{Malware Data Science} book \cite{saxe2018malware}. This dataset consists of 1,419 Windows executable files originally sourced from VirusTotal.com in 2017. It comprises of:
\begin{itemize}
    \item \textbf{Malware:} 428 Malware samples of which we could use approx. 240 samples. This represents a diverse set of malicious families prevalent during that period. All the malware samples are of type 'Backdoor'.
    \item \textbf{Benignware:} Even though the book has 991 benign ware samples consisting of legitimate software commonly uploaded by users, we were only able to use ~365 samples.
\end{itemize}
Please note that we were able to use only a subset of the samples available from the book. This limitation was necessary because the size of the decompiled executables often exceeds the input token limits of the LLMs. For example, the file \textit{Backdoor.Win32.Prorat.szb\_2874.exe} (187 KB) expands to a text file of approximately 2.5 MB upon decompilation. Consequently, because these token limits vary, the specific set of malware and benign samples used in our experiments also differed between LLMs.

\subsubsection{Contemporary Dataset}
To assess the models' ability to detect modern threats, we compiled a second dataset of recent files from 2025.
\begin{itemize}
    \item \textbf{Malware Samples}: We downloaded 120 newly submitted malware executables from Malware Bazaar (\url{malwarebazaar.com}), a public repository that provides access to active and emerging threats. The downloaded samples consists of were distributed across 49 distinct malware families. Table \ref{table:malware_family} shows the top 10 Malware families.
    \item \textbf{Benign Samples}: For our contemporary benign set, we specifically chose system drivers and other low-level utility executable. These 28 files were selected because their core function involves direct interaction with the operating system kernel and hardware—actions that often mirror the low-level manipulations performed by malware. This intentional selection of "malware-like" benign files helps create a more difficult classification problem, thereby testing the models' ability to discern malicious intent beyond superficial code similarities.
\end{itemize}

\subsection{Classification Prompt Design}

To ensure a standardized and replicable classification process, we engineered a specific zero-shot prompt (Listing. \ref{fig:prompt}) that was used for all the models. This prompt frames the task, instructs the model to act as a cyber-security expert, defines the step-by-step logic, and strictly enforces a JSON output format for parsing.

% The [!t] hint tells LaTeX to place it at the top of the page.
\begin{figure}[!ht]
    \centering
    % Use the lstlisting environment with the custom style 'promptstyle'
\begin{lstlisting}[style=promptstyle, caption={The complete prompt provided to all LLM evaluations.}, label={fig:prompt}, xrightmargin=5pt]
You are an expert in cybersecurity and reverse engineering, specializing in malware analysis. Your task is to analyze a given C code, which has been generated from decompiling an executable file using Ghidra, and determine if the file is malware.
You are free to utilize any relevant malware analysis rules or techniques, such as YARA rules, to aid in your assessment. Your final determination should be either "True" if the file is malware, or "False" if it is not.
# Step by Step instructions
1. Analyze the provided File (C code) for suspicious functions, API calls, or patterns commonly associated with malware.
2. Apply known malware analysis rules or techniques, such as YARA rules, to identify potential malicious indicators within the File.
3. Based on the analysis, determine if the File exhibits characteristics of malware.
4. Output "True" if the File is identified as malware, and "False" if it is not. Only Answer in True/False.
5. Output Format: ```{"Malware": <True/False>}```
IMPORTANT NOTE: Start directly with the output, do not output any delimiters.
Take a Deep Breath, read the instructions again, read the inputs again. Each instruction is crucial and must be executed with utmost care and attention to detail.
\end{lstlisting}
\end{figure}

\subsection{Quarantining and Licensing}

We retrieved malware samples from MalwareBazaar under its CC0 (public-domain) licensing terms, which permit free use for both commercial and non-commercial research, but disclaim any warranty or liability.

All raw binaries were stored in an encrypted, access-restricted quarantine volume, labelled by SHA256 hash and reverted after each analysis run. The analysis was carried out in isolated virtual machines with no direct internet connectivity (only via simulated, filtered network proxies). After each experiment, the virtual machine was reverted to a clean snapshot and any residual files securely wiped. Access logs and sample provenance metadata were retained; raw malware was never transferred to general-purpose systems.

\begin{table}[!htp]
\centering
\caption{Malware Family Distribution for Contemporary Dataset}
\label{table:malware_family}
\begin{tabular}{lr}
\toprule
Malware Family & Count \\
\midrule
LummaStealer & 24 \\
XWorm & 12 \\
GCleaner & 7 \\
SalatStealer & 6 \\
CoinMiner & 6 \\
AsyncRAT & 6 \\
ValleyRAT & 5 \\
SnakeKeylogger & 5 \\
Nebula & 5 \\
AgentTesla & 5 \\
\bottomrule
\end{tabular}
\end{table}
\section{Evaluation}
This section talks about the empirical evaluation of Large Language Models (LLMs) for malware classification. The analysis is presented in three stages: first, establishing a performance baseline using various standard, or "vanilla," LLMs; second, assessing the impact of fine-tuning on the best-performing model; and third, analyzing the performance drivers and limitations of these models.
\subsection{Malware Classification on Vanilla LLMs}
To establish a baseline, several state-of-the-art LLMs were evaluated on their ability to classify malware from the 2017 Baseline Dataset. The models tested include Llama 3.3 70B, Codestral, Claude 3.7 Sonnet, and Gemini 2.5 Pro. For each classification task, the model was provided with the decompiled C code from Ghidra and the corresponding static analysis of the file as context.

The performance of these models was compared against an XGBoost classifier trained on static analysis data from 1,200 samples in the baseline dataset. The XGBoost classifier achieved a near-perfect accuracy of 98.5\% on the test data, consistent with findings from similar studies \cite{salasStaticAnalysis}. This result demonstrates that a traditional machine learning model, when trained on relevant static features, can outperform vanilla LLMs on a known dataset.

Amongst the LLMs, Gemini 2.5 Pro demonstrated the highest performance. As detailed in Table II, it achieved an accuracy of 80.1\%, precision of 82.5\%, and a recall of 74.4\%. Its F1-score, a harmonic mean of precision and recall, was also the highest among the LLMs at 78.7\%, making it the strongest candidate for further fine-tuning. The other models, while showing high recall, suffered from significantly lower accuracy and precision, as shown in Table \ref{tab:2017_confusion_matrix} and Table \ref{tab:full_model_comparison_updated}.

% The \textit{Baseline Dataset} was used to create the baseline for LLMs for their ability to classify the Malware. The static analysis for each file was provided as part of the context. Table \ref{tab:2017_confusion_matrix} shows the results of the malware classification confusion matrix for Llama 3.3 70B (from Meta), Codestral (from Mistral) and Claude 3.7 Sonnet (from Anthropic) and Gemini 2.5 Pro (from Google). Table \ref{tab:full_model_comparison_updated} shows that Gemini is high performance model that can identify with 90\% accuracy. Almost all the models performed really well based Recall. We compare the performance of these LLMs with the XGBoost classifier that was trained with the static analysis data of the files from Baseline dataset with ~1200 sample data. The accuracy of the XGBoost classifier on the test data was 100\%. This XGBoost classifier performance was similar to the one observed by Salas et. al \cite{salasStaticAnalysis}. This shows that XGBoost classifier out performed the vanilla LLMs.

% \begin{table}[!ht]
% \centering
% \caption{Confusion Matrix Values for Models on 2017 Data}
% \label{tab:2017_confusion_matrix}
% \begin{tabular}{lrrrr}
% \toprule
% \textbf{Model} & \textbf{TP} & \textbf{TN} & \textbf{FP} & \textbf{FN} \\
% \midrule
% Gemini Pro 2.5 & 271 & 297 & 44 & 24 \\
% Llama 3.3 70B & 229 & 21 & 252 & 6 \\
% Codestral & 196 & 29 & 238 & 30 \\
% Claude 3.7 & 240 & 87 & 200 & 0 \\
% \bottomrule
% \end{tabular}
% \label{tab:confusion_matrices_various_models}
% \end{table}

\begin{table}[h!]
\centering
\caption{Confusion Matrix Values for Different Models on 2017 data}
\label{tab:2017_confusion_matrix}
\begin{tabular}{lcccc}
\toprule
\textbf{Model} & \textbf{TP} & \textbf{FN} & \textbf{FP} & \textbf{TN} \\
\midrule
Gemini 2.5 Pro  & 264 & 91 & 52  & 312 \\
Llama 3.3 70B   & 204 & 36 & 260 & 70  \\
Codestral       & 169 & 65 & 104 & 65  \\
Claude 3.7 Sonnet & 177 & 63 & 144 & 164 \\
\bottomrule
\end{tabular}
\end{table}

% In our experiments, we found that the Vanilla Gemini 2.5 pro had higher accuracy (89\%) and precision (86\%) than other models. The harmonic mean i.e. F1 score is also higher for the Gemini 2.5 pro model. The data comparison is show in Table \ref{tab:full_model_comparison_updated}.

\subsection{Malware classification on Fine Tuned LLM}
% Gemini 2.5 pro LLM was fine-tuned from the data sample consisting of 100 malware and 88 benign C files of de-compiled executable files, along with their static analysis with the total token count being 644348. These samples were taken from Baseline dataset. To ensure that the data were manageable and had a lower token count, we included only sample files that were less than 3 MB in size. This was supervised fine-tuning with 40 Epochs using Google Cloud Vertax AI.

% The performance of the fine tuned Gemini 2.5 Pro model on the new malware was unexpectedly great for the system setup as shown in table \ref{table:comparison_tuned}. It is able to accurately identify 100\% of actual Malware as shown in table \ref{table:gemini_comparison_final}.
% However, XGBoost classifier's accuracy drops significantly to 74\% on the Contemporary dataset.

The Gemini 2.5 Pro model was chosen for supervised fine-tuning based on its superior performance on the baseline tests. The tuning dataset consisted of 200 files from the Baseline Dataset, specifically the 100 smallest malware samples and 100 smallest benign samples, a selection method used to manage both the input token limits and cost associated for fine tuning. This fine-tuning process was conducted on Google Cloud Vertex AI over 40 epochs

The fine-tuned Gemini 2.5 Pro model was then evaluated against the Contemporary 2025 Dataset, discussed in section \ref{sec:datacollection}. The performance of the fine-tuned model was higher than even the XGBoost model, achieving 83.2\% in accuracy, 94.4\% precision, and 84.2\% recall as shown in Table \ref{table:comparison_tuned} and Table \ref{table:gemini_comparison_final}. This indicates that even though tuned LLM performed well on the contemporary test data, it still requires constant fine-tuning to understand the new evolving threats.

In contrast, the performance of the XGBoost classifier degraded significantly when applied to the new dataset, with its accuracy dropping to 74.4\%. This highlights the brittleness of traditional models, which struggle to generalize to new and evolving threats not present in their training data.

% \begin{table}[!ht]
% \centering
% \caption{Performance Metrics for Models on 2017 Data}
% \label{tab:full_model_comparison_updated}
% % \begin{tabular}{lrrrr}
% \begin{tabular}{@{}lcccc@{}}
% \toprule
% \textbf{Model} & \textbf{Accuracy} & \textbf{Precision} & \textbf{Recall} & \textbf{F1-Score} \\
% \midrule
% Gemini Pro 2.5 & \textbf{89.3\%} & \textbf{86.0\%} & 91.8\% & \textbf{88.8\%} \\
% Llama 3.3 70B & 49.2\% & 47.6\% & 97.4\% & 64.0\% \\
% Codestral & 45.6\% & 45.1\% & 86.73\% & 59.4\% \\
% Claude 3.7 & 62.0\% & 54.5\% & \textbf{100\%} & 70.6\% \\
% \midrule
% XGBoost Classifier & 98.5\% & 95.6\% & 100\% & 97.7\% \\
% \bottomrule
% \end{tabular}
% \end{table}

\subsection{Analysis on Performance of Vanilla LLMs}
% We hypothesize that an LLM's effectiveness in understanding malware code is fundamentally capped by its broader programming capabilities, causing their performance to degrade on the novel data as well. According to leading benchmarks such as Aider Polyglot \cite{aider-leaderboards} and Live Code Bench \cite{livecodebench}, even the most advanced models achieve a performance score of around 85\%. However, It is important to note that these benchmarks primarily measure an LLM's ability to generate code for novel problems, not its ability to comprehend existing code directly. Although successful code generation implies a degree of understanding, it is an indirect measure at best, suggesting the models' code comprehension abilities are similarly constrained. We further hypothesize that as the coding ability of these LLMs improves, the Malware detection without fine tuning via this technique would improve as well, making our approach one of the industry standards for malware detection.

We hypothesize that an LLM's effectiveness in malware analysis is fundamentally capped to its underlying code comprehension capabilities. Although code generation benchmarks are an indirect measure of this ability, they indicate that even top models achieve scores around 85\% (Aider Polyglot \cite{aider-leaderboards}, Live Code Bench \cite{livecodebench}), suggesting a similar ceiling for code analysis tasks without specialized training. The superior performance of the fine-tuned model demonstrates that targeted training on domain-specific data can overcome these general limitations. It is hypothesized that as the general coding and reasoning abilities of LLMs continue to advance, their baseline performance on malware detection via decompiled code will improve, potentially establishing this approach as an industry standard.

\begin{table}[!htp]
\centering
\caption{Performance Metrics for Models on 2017 Data}
\label{tab:full_model_comparison_updated}
\begin{tabular}{lrrrr}
\toprule
\textbf{Model} & \textbf{Accuracy} & \textbf{Precision} & \textbf{Recall} & \textbf{F1 } \\
\midrule
Llama 3.3 70B    & 48.1\% & 44\% & \textbf{85\%} & 57.9\% \\
Codestral & 58.1\% & 61.9\% & 72.2\% & 66.7\% \\
Claude 3.7 Sonnet& 62.2\% & 55.1\% & 73.8\% & 63.1\% \\
Gemini 2.5 Pro   & \textbf{80.1\%} & \textbf{83.5\%} & 74.4\% & \textbf{78.7\%} \\
XGBoost Classifier & 98.5\% & 95.6\% & 100\% & 97.7\% \\
\bottomrule
\end{tabular}
\end{table}

\begin{table}[!htp]
\centering
\caption{Confusion Matrix Values for Models on Contemporary 2025 Data}
\label{table:comparison_tuned}
\begin{tabular}{lrrrr}
\toprule
\textbf{Model} & \textbf{TP} & \textbf{TN} & \textbf{FP} & \textbf{FN} \\
\midrule
Gemini Pro 2.5 & 71 & 24 & 4 & 49 \\
Gemini Pro 2.5 Tuned & 101 & 22 & 6 & 19 \\
XGBoost Classifier & 106 & 24 & 4 & 14 \\
\bottomrule
\end{tabular}
\end{table}

\begin{table}[!htp]
\centering
\caption{Performance Metrics for Models on 2025 Data}
\label{table:gemini_comparison_final}
% \begin{tabular}{lrrrr}
\begin{tabular}{@{}lcccc@{}}
\toprule
\textbf{Model} & \textbf{Accuracy} & \textbf{Precision} & \textbf{Recall} & \textbf{F1} \\
\midrule
Gemini Pro 2.5 & 64.2\% & 94.7\% & 59.1\% & 72.8\% \\
Gemini Pro 2.5(T) & 83.2\% & 94.4\%	& 84.2\% & 89.0\% \\
XGBoost Classifier & 74.3\% & 81.5\% & 88.3\% & 84.8\% \\
\bottomrule
\end{tabular}
\end{table}
\section{Discussion and Future Work}

The current work demonstrates that large language models (LLMs) achieve strong performance in binary classification tasks, accurately distinguishing between benign and malicious executables.

\subsection{Contextual Information}
Integrating additional data modalities such as dynamic execution traces, sandbox detonation results, and provenance metadata into a hybrid analysis framework can elevate these results to production-grade reliability. Such systems would employ hierarchical attention and cross-modal fusion mechanisms to weigh and correlate information from diverse sources.

\subsection{Context Window}
Models with limited token windows struggle to maintain relevance across lengthy code sequences, leading to “lost-in-the-middle” failures and degraded performance on critical segments. Simply expanding context size incurs prohibitive computational and memory costs, underscoring the need for more efficient long-context processing techniques such as hierarchical chunking, adaptive attention, and memory-augmented architectures.

Promising strategies include chunk-based summarization with progressive context propagation, adaptive attention that selectively focuses on high-salience code regions, and memory-augmented modules capable of preserving long-term dependencies across analysis tasks

\subsection{Domain-specific Models}
General-purpose models offer versatile transfer learning benefits but often lack specialized understanding of malware terminology and code patterns. Domain-adaptive training methods, including continuous pretraining on curated cybersecurity datasets and parameter-efficient fine-tuning, can imbue models with necessary domain knowledge without sacrificing broader language capabilities.

Tan et al. \cite{tan2024llm4decompile} describe methods to use LLMs to decompile binaries. Future work can replace the use of Ghidra with such models to evaluate the efficacy of techniques that use LLMs end to end in the malware detection pipeline.

\subsection{Explainability}
Transparency remains a core concern for LLM-based security systems. The black-box nature of these models hinders analyst trust and regulatory compliance, as interpretable explanations of classification decisions are essential for validation and debugging.
% \section{Future Work}

% Future research can focus on developing multi-modal hybrid architectures that can seamlessly integrate static code features with dynamic execution traces and external metadata. Such systems would employ hierarchical attention and cross-modal fusion mechanisms to weigh and correlate information from diverse sources.

\bibliographystyle{unsrt}
\bibliography{main}

\end{document}